# Nonlinear atomic tunnelling boosted by bright squeezed vacuum

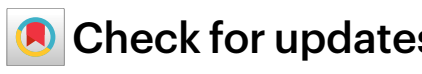

Zhejun Jiang[1,4], Shengzhe Pan[1,4], Jianqi Chen[1], Mingyu Zhu[1], Chenhao Zhao[1], Yiwen Wang[1], Ru Zhang[1], Jianshi Lu[1], Lulu Han[1], Suwen Xiong[1], Dian Wu[1], Wenxue Li[1], Shicheng Jiang[1]✉, Hongcheng Ni[1]✉ & Jian Wu[1,2,3]✉

Nonlinear optical processes, mediated by multiphoton interactions rather than single-photon response, are routinely exploited to enable a range of light-based functionalities in devices and applications. Nonlinear effects are enhanced by higher-intensity fields, which is a limiting strategy owing to potential radiation damage. An alternative strategy relies on the fluctuation redistribution typical of quantum light[1–4], but experimental demonstrations at the most fundamental level have been limited. Here we report experimental nonlinear tunnelling ionization of isolated atoms, a pivotal nonlinear process that drives high-harmonic generation and forms the basis of attosecond science, boosted by quantum light—bright squeezed vacuum (BSV). A BSV light with an average pulse energy of 300 nJ achieves an effective intensity equivalent to that of a coherent light with 7.1 μJ, demonstrating a more than 20-fold quantum boost in the nonlinear effect from BSV light. This boost is revealed by matching the peaks of the photoelectron momentum spectra produced by the BSV and coherent light as measured by angular streaking. Furthermore, we demonstrate control of the effective intensity of the BSV by tuning the correlation function at fixed average pulse energy, establishing a robust method to tailor nonlinear processes by quantum statistics rather than classical intensity scaling. These findings may facilitate the development of quantum-controlled strong-field dynamics using tailored quantum light sources.

Nonlinear light–matter interactions driven by multiple photons enable revolutionary abilities beyond perturbative optics, in contrast to the well-known single-photon process. At the heart of these laser-driven nonlinear processes lies tunnelling ionization[5–12], which forms the cornerstone of ultrafast science. It not only drives femtochemistry[13–16] but also serves as the vital first step in high-harmonic generation (HHG)[17–21]—the primary mechanism for producing attosecond pulses[22–24]—thereby enabling the direct exploration of electron dynamics on its natural time scale[25–31]. However, conventional approaches using coherent light enhance tunnelling ionization through brute-force intensity scaling face an inherent limitation: material damage threshold constrains maximum accessible intensity of the laser field. Quantum light, for example, bright squeezed light, breaks this pattern by leveraging intrinsic photon correlations and enhanced amplitude fluctuation[1,2]. These quantum properties enable a superior nonlinear boost at low excitation powers, accessing non-classical interaction regimes impossible with coherent light[32–34].

Recently, bright squeezed vacuum (BSV) quantum light sources have been generated by high-gain parametric down-conversion in nonlinear crystals[35–38]. Characterized by non-classical photon number statistics, the BSV light has enabled several breakthroughs, including the observation of super-Poissonian statistics in high-energy photons through HHG in crystals[1,4,39] and photoelectrons by multiphoton ionization from nanotips[3,40], as well as enhanced efficiency and spectral broadening in these nonlinear processes[1,2,4,39,41–50]. Despite these remarkable advances, the fundamental quantum dynamics of BSV-driven tunnelling ionization in the most basic isolated atomic systems free from collective effects remain unexplored, particularly because rare-gas atoms have relatively high ionization potentials that challenge the achievable BSV intensities now. A direct, intensity-calibrated signature of quantum statistics imprinted on the photoelectron energy spectrum of an isolated atom and a means to effectively control it remain unknown.

In this work, we report quantum-boosted nonlinear tunnelling of an isolated sodium atom driven by a femtosecond BSV pulse. The use of sodium, with its low ionization potential, is important to overcome the limitation of the achievable intensity of the BSV light, while it serves as an elementary atomic system in which material-specific effects are absent. Our experiment shows a striking quantum advantage: a BSV light with 300 nJ average pulse energy produces equivalent peak photoelectron momentum to that generated by 7.1 μJ of coherent light, representing a more than 20-fold boost in the nonlinear effect through the engineering of quantum statistical properties. This boost is revealed through angular streaking[51], which maps the instantaneous vector potential at the moment of ionization onto the final photoelectron momentum. Furthermore, we develop an approach to actively control the effective intensity of the BSV light to boost the nonlinear process by tuning the correlation function while maintaining fixed average pulse energy. The quantum-boosted tunnelling ionization is well supported by a quantum Ammosov–Delone–Krainov (QADK) theory, developed here to account for light–electron entanglement that imprints quantum statistical property of the BSV light onto emitted electrons. These results establish quantum light as a powerful tool for strong-field physics, offering both an insightful understanding of non-classical light–matter interactions and practical methods for controlling extreme nonlinear processes.

[1]State Key Laboratory of Precision Spectroscopy, East China Normal University, Shanghai, China. [2]Collaborative Innovation Center of Extreme Optics, Shanxi University, Taiyuan, China. [3]Chongqing Key Laboratory of Precision Optics, Chongqing Institute of East China Normal University, Chongqing, China. [4]These authors contributed equally: Zhejun Jiang, Shengzhe Pan. ✉e-mail: scjiang@lps.ecnu.edu.cn; hcni@lps.ecnu.edu.cn; jwu@phy.ecnu.edu.cn

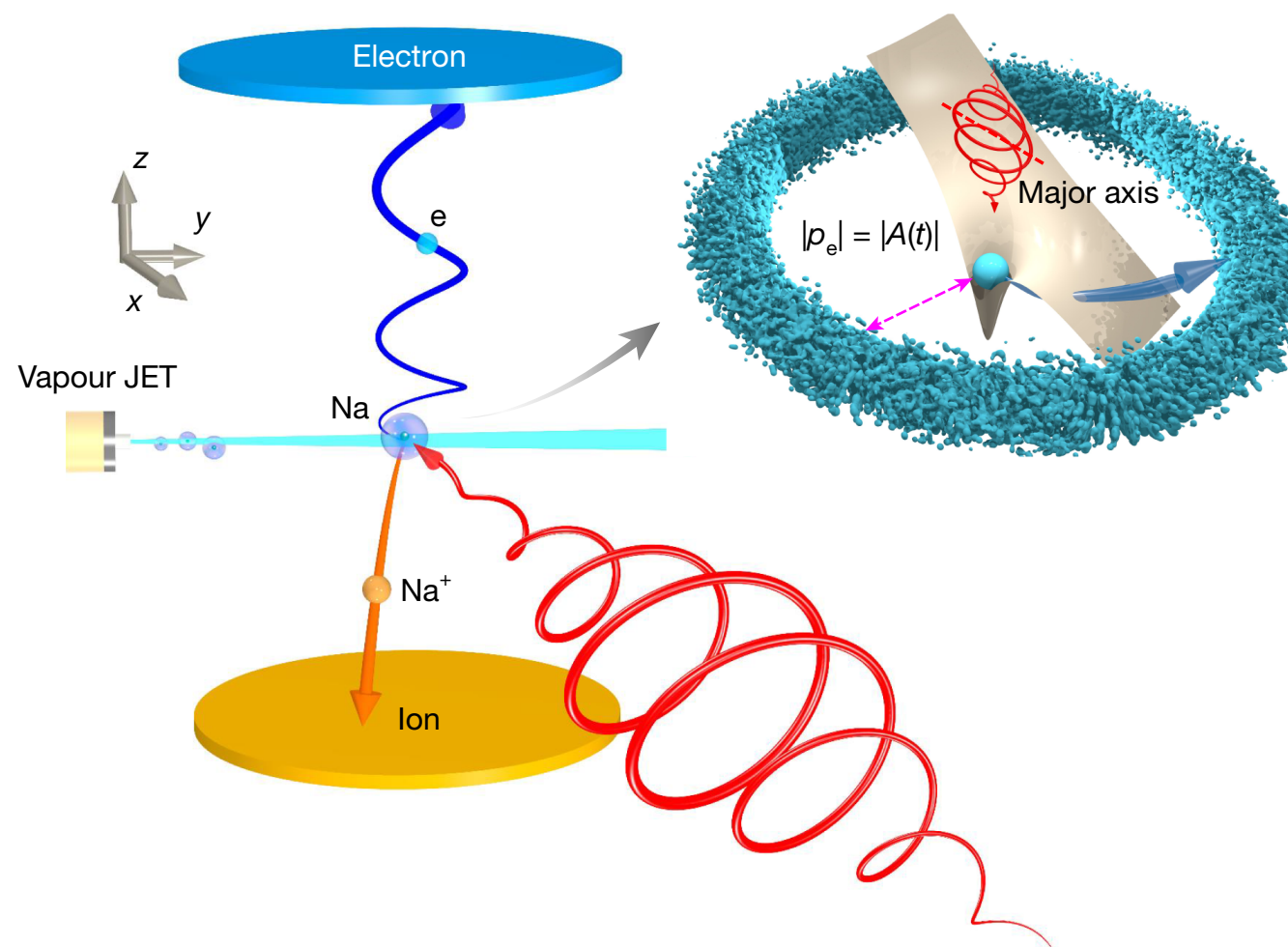

**Fig. 1 | Experimental scheme.** The schematic illustrates the tunnelling ionization of sodium atoms from a vapour jet induced by elliptically polarized coherent light or quantum BSV sources. Ionized electrons and parent ions are detected in coincidence. The inset shows the angular streaking principle, in which an electron tunnelling through the laser-suppressed atomic potential acquires a final momentum determined by the instantaneous vector potential and a weight determined by the effective intensity of the elliptically polarized pulse at the ionization instant.

## Experimental setup

To demonstrate the quantum-boosted nonlinear tunnelling of sodium atoms, we perform comparable measurements driven by either (1) a conventional laser pulse in a classical coherent state or (2) a quantum BSV laser pulse. The coherent light source with a pulse duration of 70 fs and a central wavelength of 1,580 nm is generated by a commercial TOPAS-Prime system pumped by a multipass Ti:sapphire amplifier at a 10 kHz repetition rate. By contrast, the BSV light source is generated by high-gain parametric down-conversion in two cascaded 3-mm type-I β-barium borate (BBO) crystals[35,36], pumped by the same amplifier with a pulse duration of 28 fs and a central wavelength of 790 nm. The resulting BSV light has a duration of 150 fs and exhibits a broad spectral bandwidth spanning 1,400–1,800 nm, centred at 1,580 nm to ensure comparability with the coherent light source. The peak intensity of the coherent light in the interaction region is estimated to be $1\times10^{13}$ W cm$^{-2}$. To facilitate angular streaking, the ellipticity of both coherent and BSV lights is set at 0.7.

Both light sources are tightly focused onto a dilute sodium vapour jet using a silver-coated concave mirror within an ultrahigh-vacuum chamber of a cold-target recoil ion momentum spectrometer[52,53], as shown in Fig. 1. The ionization potential of the sodium atom is 5.14 eV. The generated photoelectrons and photoions are detected by two time- and position-sensitive microchannel plate detectors positioned at opposite ends of the spectrometer. We select photoelectron events in the single ionization channel by applying a coincidence condition of $|p_z^{(\mathrm{ele})} + p_z^{(\mathrm{ion})}| < 0.2$ a.u. along the time-of-flight axis of the spectrometer with the best momentum resolution. Given the low tunnelling ionization probability of about 0.01% per laser pulse, we accumulate coincidence events over an extended duration and rebin the recorded electron counts to reflect the statistical fluctuation of the ionization rate. For more details of the experimental setup, please refer to the Methods.

## Quantum-boosted tunnelling

Figure 2a presents the probability distributions of emitted electron number driven by coherent and BSV pulses while maintaining the same average electron number. The electron number statistics show a fundamental distinction between coherent and BSV-driven ionization. For coherent light excitation (Fig. 2a, blue circles), the electron number distribution following a Poissonian profile (Fig. 2a, blue line) confirms that the strong-field nonlinearity in the tunnelling regime preserves the Poissonian character of photon number statistics, as expected for coherent processes[3]. By contrast, BSV-driven ionization (Fig. 2a, orange squares) exhibits pronounced non-Poissonian statistics with a much broader distribution, directly inheriting the non-classical photon number statistics of the BSV light, which is successfully reproduced by our theoretical model (Fig. 2a, orange line; see the Methods for details).

Beyond non-classical electron-number statistics, the quantum BSV light provides an advantage of boosting nonlinear response with markedly extended spectral characteristics. The extension of photoelectron energy spectra for isolated atomic targets has been theoretically predicted[54–56] but not verified in experiments. Through angular streaking, we directly characterize this advantage in a quantitative manner: the boosted nonlinearity manifests as extended tails in the photoelectron kinetic energy spectrum driven by a BSV pulse compared with that driven by a coherent light with the same effective intensity. To achieve this, we generate both coherent and BSV light sources with matched ellipticity and effective intensity. As shown in Fig. 1, the angular streaking technique driven by an elliptically polarized pulse allows mapping the vector potential $A_0$ at the ionization instant to the momentum magnitude of the released photoelectron $p_e$, that is, $\langle p_e \rangle = A_0$, as predicted by the strong-field approximation. This relation arises because the ionized electron is primarily driven by the laser field after tunnelling. For the BSV light, which exhibits inherent field strength fluctuations averaging to a vanishing mean field, we operationally define its effective intensity as that of a coherent pulse that produces the same most probable photoelectron momentum (Fig. 2b, magenta arrow) under identical angular streaking conditions, thereby enabling a direct comparison. Note that the actual intensity is related to the average pulse energy, whereas the effective intensity is imprinted on the peak momentum of the photoelectron as calibrated by angular streaking. This configuration requires only 300 nJ of the BSV light with a second-order correlation function $g^{(2)}(0) = 1.5$ (abbreviated as $g^{(2)}$ hereafter) to achieve equivalent strong-field effects as 7.1 μJ of the coherent light, demonstrating a more than 20-fold enhancement in the nonlinear effect. Crucially, we confirm that a 300 nJ coherent light produces no measurable tunnelling, unambiguously proving that the quantum statistical properties of BSV drive the enhanced nonlinear response rather than classical intensity effects. The 20-fold quantum enhancement observed here refers to the fact that the effective intensity of the BSV field matches that of a coherent light with 20 times as much pulse energy, rather than an increase in the total ionization yield.

The BSV-driven ionization shows a striking quantum signature: a significantly broadened photoelectron kinetic energy spectrum compared

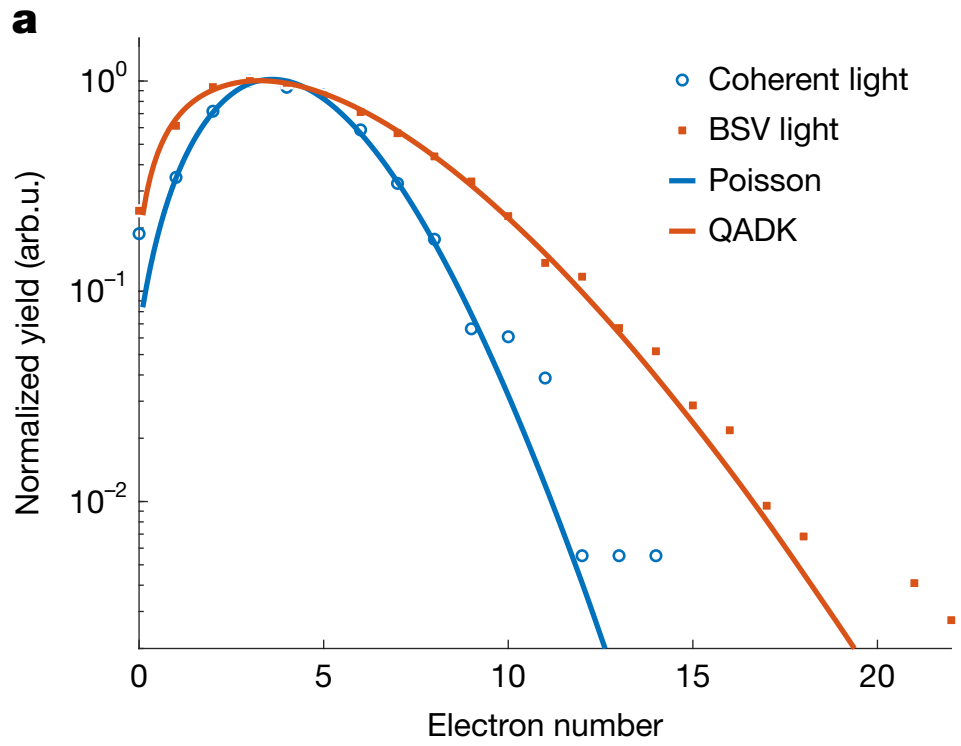

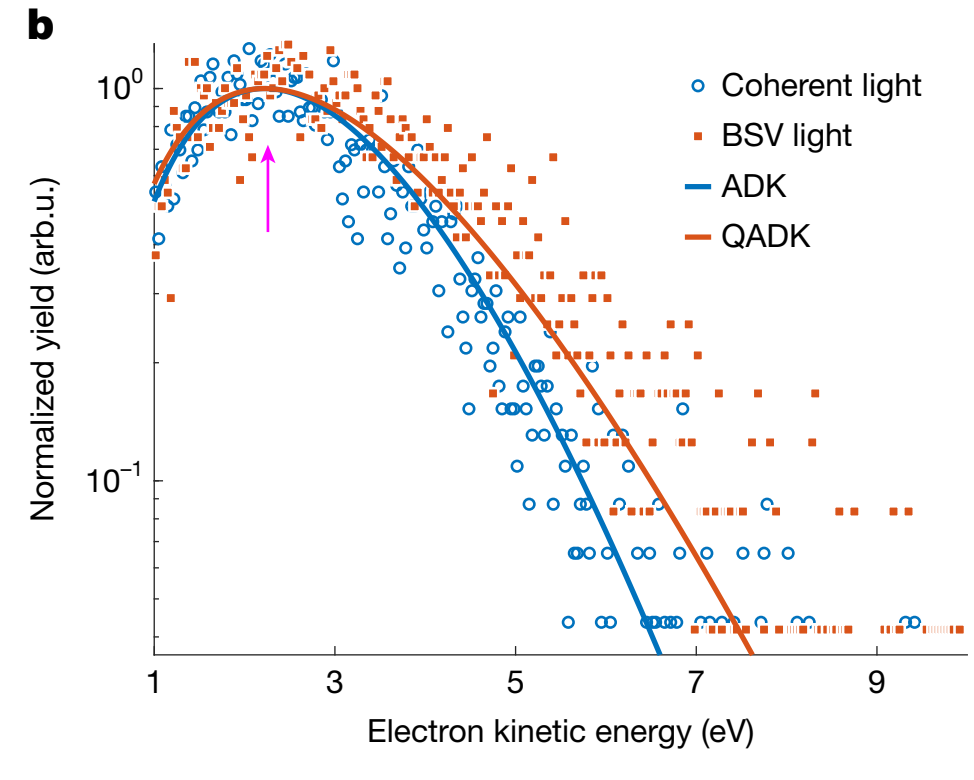

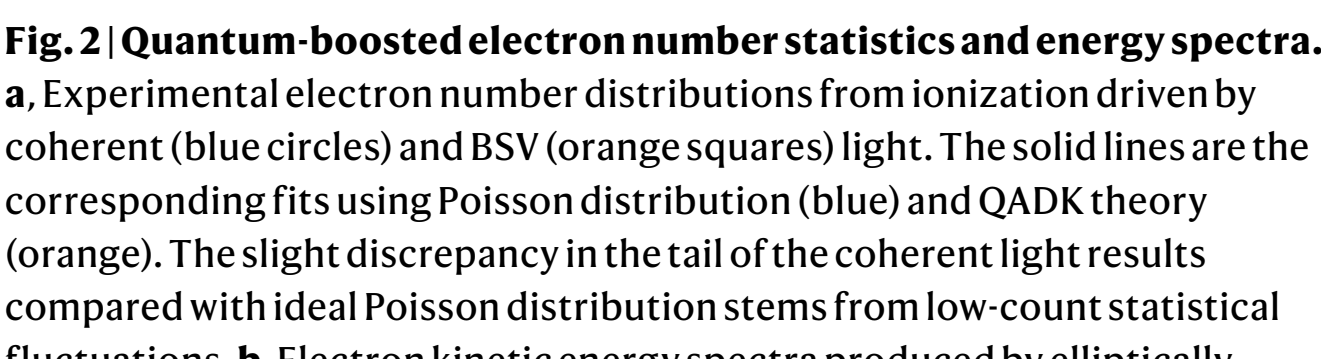
**Fig. 2 | Quantum-boosted electron number statistics and energy spectra.** **a**, Experimental electron number distributions from ionization driven by coherent (blue circles) and BSV (orange squares) light. The solid lines are the corresponding fits using Poisson distribution (blue) and QADK theory (orange). The slight discrepancy in the tail of the coherent light results compared with ideal Poisson distribution stems from low-count statistical fluctuations. **b**, Electron kinetic energy spectra produced by elliptically polarized coherent (blue circles) and BSV (orange squares) light. The solid lines are simulated by semiclassical ADK (blue) and QADK (orange) theories. The matched peak position (magenta arrow) confirms equal effective intensities, whereas the extended high-energy tail under BSV directly reflects its amplitude-stretched quantum fluctuations. The striated structure is an artefact of limited counting statistics, common in discrete-event detection experiments[3,39]. arb.u., arbitrary units.

with the classical case (Fig. 2b). Although both light sources generate peaks at equivalent electron momentum, maintaining matched effective intensities, the BSV-driven spectrum (Fig. 2b, orange squares) shows a pronounced high-energy tail absent in the spectrum driven by coherent light (Fig. 2b, blue circles). This spectral broadening provides direct experimental evidence of enhanced energy transfer in strong-field light–atom interactions mediated by quantum correlations.

To rigorously model the quantum enhancement, we have developed a QADK theory based on an entangled light–electron Hamiltonian (see the Methods for details). This framework shows that the quantum statistical nature of the BSV light redistributes the ionization probability across a broad range of final electron momenta, directly explaining the enhanced nonlinear effect and spectral broadening we observe. In Fig. 2b, the solid lines compare theoretical predictions from the semiclassical ADK[57–59] and QADK theories with the experimental results. The semiclassical ADK theory accurately reproduces the narrow spectrum produced by the coherent light source. By contrast, the QADK theory, which incorporates the multi-mode quantum statistics ($N = 5$ Schmidt modes[60]), provides good agreement with the broadened spectrum measured under the BSV light.

## Quantum tunability

Furthermore, we introduce a method for controlling strong-field processes by actively tuning the quantum correlation function $g^{(2)}$ of the BSV light while maintaining constant average pulse energy. In our experiment, we first increase the pump power entering the nonlinear crystal and then attenuate the average power of the generated BSV back to the original level. Physically, this procedure simultaneously increases the mode number $N$ (ref. 60) and decreases the squeezing parameter $r$ accordingly under the constraint of fixed average pulse energy, which in turn leads to the variation of the $g^{(2)}$ parameter (see the Methods for details). Figure 3a demonstrates this principle through photoelectron kinetic energy spectra generated by BSV light with varying $g^{(2)}$. As $g^{(2)}$ increases from 1.00 to 1.39, the spectra systematically shift towards higher energies despite the use of identical average pulse energy. The peaks of the spectra can be extracted and converted into effective intensities (see Methods for details). The extracted spectral peaks show a linear scaling between $g^{(2)}$ and effective intensity $I_{eff}$ (Fig. 3b), establishing three key advances: (1) precise effective intensity control at constant average pulse energy; (2) a direct connection between quantum statistics and nonlinear enhancement;

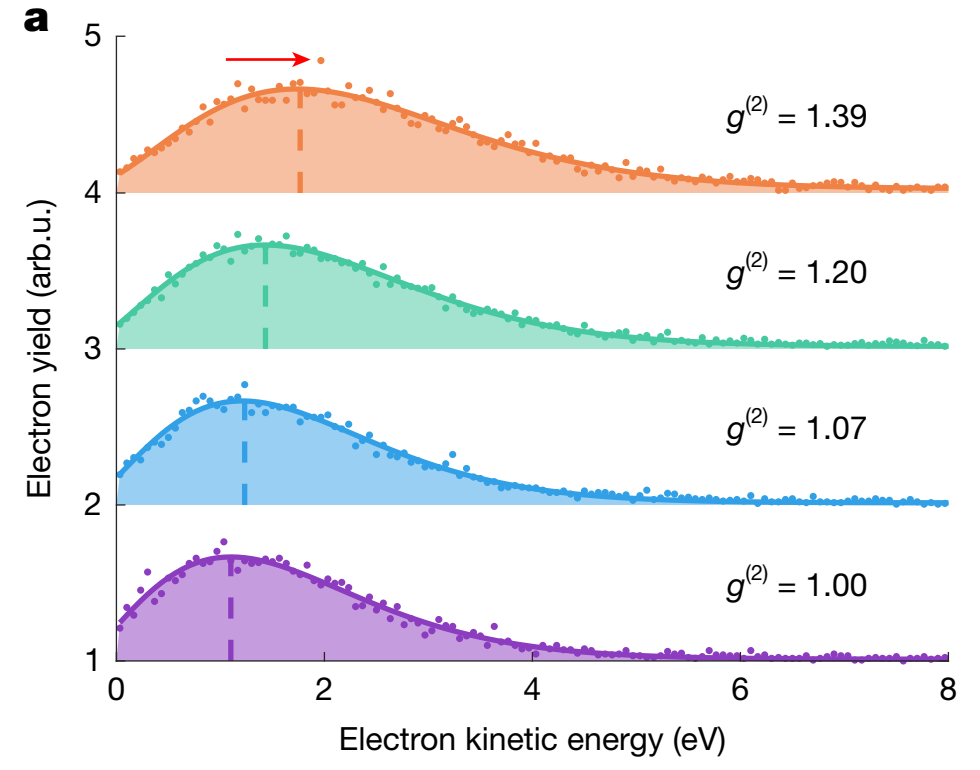

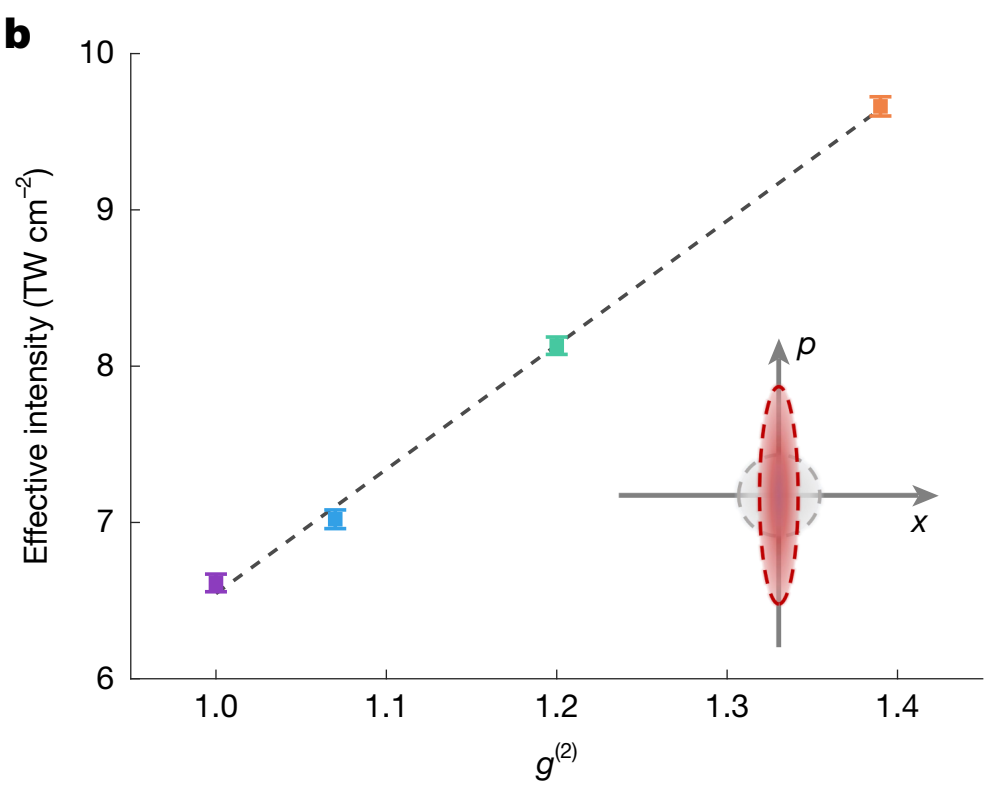

**Fig. 3 | Quantum control of photoelectron spectra by tuning $g^{(2)}$.** **a**, Electron kinetic energy spectra resulting from BSV light with varying second-order correlation function $g^{(2)}$. The systematic shift towards higher energies with increasing $g^{(2)}$ demonstrates enhanced nonlinear interaction strength through quantum correlation engineering. **b**, Linear scaling between effective intensity and $g^{(2)}$, establishing quantum statistics as a control parameter for strong-field processes. The inset shows the quantum statistical nature of BSV light, in which reduced phase fluctuations and enhanced amplitude fluctuations enable the observed intensity enhancement at constant average pulse energy. arb.u., arbitrary units.

and (3) new abilities for optimizing quantum-enhanced strong-field phenomena. As detailed in the Methods, for BSV, this linear scaling law follows

$$I_{\mathrm{eff}} \propto P[g^{(2)}-1], \tag{1}$$

obtained under the assumption of a finite number of equally squeezed, independent Schmidt modes[60], where $P$ represents the constant average pulse energy. This expression establishes $g^{(2)}$ as a powerful experimental knob for tailoring quantum light–matter interactions, opening possibilities for intensity-tunable attosecond sources and optimized nonlinear spectroscopy at fixed average pulse energy of the driving quantum light.

## Discussion

In summary, our work, as an experimental demonstration of atomic tunnelling driven by quantum light, establishes BSV as a transformative tool for strong-field physics through the following fundamental advances. First, the implementation based on isolated sodium atoms conclusively pins down the origin of ionization enhancement and spectral broadening to quantum statistical properties of the BSV light. Second, by using the angular streaking technique, we have performed a direct and precise calibration of the effective intensity of BSV light, thereby exhibiting a more than 20-fold quantum enhancement in the nonlinear tunnelling effect, in which identical peak electron momentum is achieved with 300 nJ BSV pulses compared with 7.1 μJ coherent pulses. Third, we introduce a new quantum control method in which the effective intensity can be precisely tuned by the correlation function while maintaining constant average pulse energy. Fourth, we have developed a QADK theory that incorporates light–electron entanglement, revealing the quantum statistical nature of the BSV field as the mechanism behind the enhanced and broadened photoelectron momentum distribution. These advances, spanning elementary quantum signature characterization, precise effective intensity calibration, active control and theoretical verification, establish a comprehensive framework for quantum-enhanced strong-field phenomena, opening new avenues for efficient attosecond sources, controlled electron dynamics and tailored nonlinear spectroscopy using engineered quantum light. Our results not only advance fundamental understanding of quantum light–matter interactions but also position strong-field quantum optics as a promising frontier for exploring extreme nonlinear processes with unprecedented control and efficiency.

## Online content

Any methods, additional references, Nature Portfolio reporting summaries, source data, extended data, supplementary information, acknowledgements, peer review information; details of author contributions and competing interests; and statements of data and code availability are available at https://doi.org/10.1038/s41586-026-10485-9.

# Methods

### Experimental details

Both coherent and quantum light sources are pumped with the same femtosecond laser pulse (790 nm, 28 fs, 10 kHz) generated by a Ti:sapphire multipass amplifier laser system (Femtolaser). The coherent light source centred at 1,580 nm and with a pulse duration of 70 fs, is produced by a commercial optical parametric amplifier (Light Conversion, TOPAS-Prime). For the quantum BSV light source, the pump beam collimated to a 4-mm diameter is propagated through two cascaded 3-mm BBO crystals. Both BBO crystals are cut for type-I collinear frequency-degenerate phase matching to generate high-gain parametric down-conversion. Here, the optical axes are oriented oppositely in the horizontal plane to minimize the spatial walk-off. The distance between two crystals is set at 80 cm so that only the spatial mode with the lowest diffraction undergoes the phase-sensitive amplification. The pulse duration of the BSV light is measured to be approximately 150 fs using the technique of cross-correlation frequency-resolved optical gating based on sum-frequency generation of the to-be-calibrated BSV pulse and a reference infrared pulse. The effect of different pulse durations for coherent and BSV lights is discussed in the section 'Effect of pulse duration'.

The second-order correlation function $g^{(2)}$ of the generated BSV is measured using the standard Hanbury Brown–Twiss technique. The BSV source is split at a nonpolarizing 50/50 beam splitter into two channels, the photon number distributions of which are measured with two fast InGaAs photodiodes. The signals from the photodiodes are recorded by a multichannel oscilloscope and then integrated over time to present the total photon numbers. Finally, the second-order correlation function is calculated by $g^{(2)} = \langle n_1 n_2 \rangle / (\langle n_1 \rangle \langle n_2 \rangle)$, where $n_1$ and $n_2$ are photon numbers measured for two channels. Considering the typical measured $g^{(2)}$ value of 1.5, there are multiple frequency modes amplified in the BSV generation process, which has been further confirmed by a spectral filtering of the BSV light using a standard 4-$f$ monochromator (see section 'Mode analysis of photon number statistics' for details). It is worth noting that to preserve the peak intensity required for tunnelling ionization, no spectral filtering is applied in our photoionization experiments. The details of controlling the correlation function $g^{(2)}$ and validating the multi-mode BSV light are demonstrated in the section 'Mode analysis of photon number statistics'.

The beam diameters for both the coherent light and the BSV light are approximately 3 mm and are tightly focused by a concave silver mirror (focal length $f$ = 75 mm) inside an ultrahigh-vacuum chamber onto the sodium vapour jet. The sodium vapour is produced in a resistively heated crucible at 170 °C and collimated by a 2-mm skimmer. The peak intensity of the coherent light in the interaction region is estimated to be $1 \times 10^{13}$ W cm$^{-2}$. The corresponding Keldysh parameter is calculated to be 1.48, which places the ionization process in the non-adiabatic tunnelling regime. The photoelectron energy spectrum (Fig. 2b) featuring a smooth structure and the absence of resonant peaks[61] also indicates that ionization proceeds through a tunnelling pathway, without significant population of intermediate states such as Na(4$s$). This distinguishes our results from regimes in which atomic saturation and resonances suppress quantum-optical enhancement effects[62] and allows for an observation of the BSV-induced enhancement.

The strong-field tunnelling ionization of a single atom generates both photoelectron and photoion, whose momenta are coincidently measured using the cold-target recoil ion momentum spectroscopy reaction microscope[52,53]. These charged particles are accelerated by a homogeneous electric field (7 V cm$^{-1}$) with the assistance of a weak magnetic field (11 G) and finally strike the detectors at the opposite ends of the spectrometer. The momenta of the emitted electron $\mathbf{p}^{(\mathrm{ele})}$ and the corresponding parent ion $\mathbf{p}^{(\mathrm{ion})}$ for each laser–atom interaction event are reconstructed from the measured time of flight and position of impact during the offline analysis. The principle of momentum conservation in the centre-of-mass frame dictates that the momentum sum of the electron and ion from the same interacting atom should vanish, that is, $\mathbf{p}^{(\mathrm{ele})} + \mathbf{p}^{(\mathrm{ion})} = \mathbf{0}$. In practice, the coincidence condition of $|p_z^{(\mathrm{ele})} + p_z^{(\mathrm{ion})}| < 0.2$ a.u. along the time-of-flight axis of the spectrometer, which has the highest momentum resolution, is used to select the right coincidence events from other background signals or false coincidences.

Both coherent and quantum light sources are converted from linear to elliptical polarization by a broadband quarter-wave plate. By rotating the fast axis of the plate relative to the initial linear polarization, the relative intensity between the two orthogonal components is controlled while introducing a fixed π/2 phase shift, thereby producing the desired ellipticity of 0.7, which is appropriate for performing angular streaking analysis using the accessible BSV light (see section 'Choice of elliptical polarization' for details).

### QADK theory

The semiclassical ADK theory[57–59], commonly used in strong-field ionization, treats the electron quantum mechanically while describing the light as a classical field. It represents the adiabatic limit of the strong-field approximation[12,63,64] and the Perelomov–Popov–Terent'ev theory[65–68]. In this framework, the doubly differential ionization rate is given by

$$\Gamma_{\mathrm{ADK}}(p_{\mathrm{e}}, t) = F(t)\exp\left\{-\frac{2[2I_{\mathrm{p}} + (p_{\mathrm{e}} - A(t))^2]^{3/2}}{3F(t)}\right\}, \tag{2}$$

where Coulomb-related prefactors are not included here for consistency. For circular polarization, the time dependence factorizes out, allowing us to directly replace $A(t)$ with $A_0$ and $F(t)$ with $F_0$. For elliptical polarization, we replace $A(t)$ with $\varepsilon A_0$ and $F(t)$ with $F_0$ for ionization along the major axis or replace $A(t)$ with $A_0$ and $F(t)$ with $\varepsilon F_0$ for that along the minor axis. For simplicity of the expression, we adopt the substitution $A(t) \to A_0$ and $F(t) \to F_0$, and the singly differential rate is expressed as

$$\Gamma_{\mathrm{ADK}}(p_{\mathrm{e}}) = F_0\exp\left\{-\frac{2[2I_{\mathrm{p}} + (p_{\mathrm{e}} - A_0)^2]^{3/2}}{3F_0}\right\}. \tag{3}$$

This expression is often further expanded for small initial transverse momentum, leading to an alternative form

$$\Gamma_{\mathrm{ADK}}(p_{\mathrm{e}}) = F_0\exp\left\{-\frac{2(2I_{\mathrm{p}})^{3/2}}{3F_0}\right\}\exp\left\{-\frac{\sqrt{2I_{\mathrm{p}}}}{F_0}(p_{\mathrm{e}} - A_0)^2\right\}.$$

The quantum enhancement observed in our experiment is fundamentally rooted in the quantum statistical property of the BSV light, which is imprinted onto the emitted electron by light–electron entanglement inherent to the quantum-optical description of the tunnelling ionization process. To rigorously capture this physics, we have developed a QADK theory based on the entangled light–electron Hamiltonian, given by

$$\widehat{H} = \underbrace{\widehat{\mathbf{p}}^2/2 + V(\hat{\mathbf{r}})}_{\widehat{H}_{\mathrm{A}}} + \underbrace{\omega\widehat{a}^{\dagger}\widehat{a}}_{\widehat{H}_{\mathrm{F}}} + \underbrace{\widehat{\mathbf{p}}\cdot\widehat{\mathbf{A}} + \widehat{\mathbf{A}}^2/2}_{\widehat{H}_{\mathrm{int}}}, \tag{4}$$

where $\widehat{\mathbf{A}} = \frac{g}{\omega}(\widehat{a}\boldsymbol{\varepsilon} + \widehat{a}^{\dagger}\boldsymbol{\varepsilon}^*)$ is the vector potential operator, $\boldsymbol{\varepsilon}$ is the polarization vector and $g$ is the coupling strength. The electron is initially in the ground state $|g\rangle$, and the light is initially in a BSV state $\widehat{S}|0\rangle$, with $\widehat{S} = \exp\left(\frac{1}{2}\xi^*\widehat{a}^2 - \frac{1}{2}\xi\widehat{a}^{\dagger 2}\right)$ the squeezing operator, where the squeezing parameter $\xi = re^{\mathrm{i}\phi}$, with $r$ and $\phi$ the squeezing amplitude and squeezing phase, respectively. The initial light–electron state is, therefore, $|\Psi_0\rangle = |g\rangle \otimes \widehat{S}(\xi)|0\rangle$, which will evolve into an entangled light–electron state as their interaction initiates.

To facilitate calculations of entangled light–electron dynamics, we apply a unitary transformation $\widetilde{\widetilde{O}} = \widehat{S}^\dagger \widehat{U}_F^\dagger \widehat{O} \widehat{U}_F \widehat{S}$ with $\widehat{U}_F = \exp(-i\widehat{H}_F t)$, which turns the Hamiltonian into

$$\widetilde{\widetilde{H}} = \widehat{H}_A + \underbrace{\widehat{\mathbf{p}} \cdot \widehat{\mathbf{A}}_S + \widehat{\mathbf{A}}_S^2/2}_{\widetilde{\widetilde{H}}_{\mathrm{int}}}, \tag{5}$$

where the transformed vector potential $\widehat{\mathbf{A}}_S = \widehat{S}^\dagger \widehat{U}_F^\dagger \widehat{\mathbf{A}} \widehat{U}_F \widehat{S}$, and the initial state becomes $|\widetilde{\Psi}_0\rangle = |g\rangle \otimes |0\rangle$, so that the light part of the initial state is simply vacuum in the transformed gauge.

For a BSV state that has a large squeezing amplitude $r$, $\widehat{\mathbf{A}}_S$ can be approximated by

$$\widehat{\mathbf{A}}_S = -iA_0(\widehat{a} - \widehat{a}^\dagger)\boldsymbol{\varepsilon}(t) = \underbrace{A_0\boldsymbol{\varepsilon}(t)}_{\text{classical}} \cdot \underbrace{\sqrt{2}\widehat{p}_\Lambda}_{\text{quantum}} \equiv \widehat{A}_{\mathrm{eff}}\boldsymbol{\varepsilon}(t), \tag{6}$$

where $\widehat{p}_\Lambda = (\widehat{a} - \widehat{a}^\dagger)/\sqrt{2}i$ is the momentum quadrature of BSV light, $\boldsymbol{\varepsilon}(t) = \frac{i}{2}(e^{-i\omega t}\boldsymbol{\varepsilon} - e^{i\omega t}\boldsymbol{\varepsilon}^*)$ with $\boldsymbol{\varepsilon} = \mathbf{e}_x + i\varepsilon\mathbf{e}_y$, $A_0 \equiv 2\frac{g}{\omega}\sqrt{\frac{n}{1+\varepsilon^2}}$ with $n$ the average photon number is the vector potential amplitude defined in such a way that it matches its classical counterpart with the same actual intensity because $I_{\mathrm{BSV}} = \frac{c\varepsilon_0}{\omega^2}\langle 0|\widehat{\mathbf{A}}_S^2|0\rangle = \frac{c\varepsilon_0}{2\omega^2}(1+\varepsilon^2)A_0^2 = I_{\mathrm{coh}}$ with $\varepsilon_0$ the vacuum permittivity and $c$ the vacuum light speed, and $\widehat{A}_{\mathrm{eff}} \equiv \sqrt{2}A_0\widehat{p}_\Lambda$ is the effective vector potential under BSV quantum light. Correspondingly, the effective electric field $\widehat{F}_{\mathrm{eff}} \equiv \sqrt{2}F_0\widehat{p}_\Lambda$. We note that the stretched quadrature has been rotated to the momentum direction in equation (6). In this transformed gauge, the classical and quantum aspects of the light–electron interaction factorize, which greatly simplifies the calculation of the entangled dynamics. In particular, the quantum-optical description proceeds in close analogy to its classical counterpart: we replace the classical vector potential $A_0$ with the effective vector potential $\widehat{A}_{\mathrm{eff}}$ and the classical electric field $F_0$ with $\widehat{F}_{\mathrm{eff}}$, whereas all other elements of the theory remain unchanged. It is important to note, however, that the light momentum $\widehat{p}_\Lambda$ enters the effective fields $\widehat{A}_{\mathrm{eff}}$ and $\widehat{F}_{\mathrm{eff}}$, and the distribution of $p_\Lambda$ encodes the information of the quantum statistical property of the quantum light. For BSV, whose initial state is vacuum $|0\rangle$ in the transformed gauge, it can be expanded in the light momentum basis:

$$|0\rangle = \int dp_\Lambda |p_\Lambda\rangle\langle p_\Lambda|0\rangle = \pi^{-1/4}\int dp_\Lambda e^{-p_\Lambda^2/2}|p_\Lambda\rangle. \tag{7}$$

Therefore, the quantum statistical weight of light momentum $p_\Lambda$ corresponds to $\chi(p_\Lambda) = e^{-p_\Lambda^2}$ in the normalized probability distribution. The initial vacuum state $|0\rangle$ is expanded in the momentum quadrature basis $p_\Lambda$ because $\widehat{\mathbf{A}}_S$ is diagonal in this representation. This simplification relies on the fact that, under large squeezing, the contribution from the conjugate position quadrature becomes negligible. This property allows us to reduce the dimensionality of the expansion. By contrast, other studies expand the initial state in a coherent-state basis[2,40] or a number-state basis[69]. Although all expansions are, in principle, equivalent, it is important to note that the coherent-state basis is not orthogonal; therefore, we cannot ignore off-diagonal elements and expect to obtain identical results.

In this framework, the doubly differential ionization rate under a single-mode BSV light is given by

$$\Gamma_{\mathrm{QADK}}(p_e, p_\Lambda) \propto \underbrace{\sqrt{2}F_0 p_\Lambda \exp\left\{-\frac{2[2I_p + (p_e - \sqrt{2}A_0 p_\Lambda)^2]^{3/2}}{3\sqrt{2}F_0 p_\Lambda}\right\}}_{\gamma(p_e, p_\Lambda)} \underbrace{e^{-p_\Lambda^2}}_{\chi(p_\Lambda)}, \tag{8}$$

where $\chi(p_\Lambda) = e^{-p_\Lambda^2}$ stems from the quantum statistical property of the BSV light. The QADK theory is a generalization of the semiclassical ADK theory to the quantum domain. It explicitly accounts for the entanglement between the photoelectron momentum $p_e$ and the light momentum $p_\Lambda$, which imprints the quantum statistical property of BSV on the emitted electrons.

The derivation above for the single-mode BSV can be extended to the multi-mode case[60] by replacing the field momentum $p_\Lambda$ by $P_\Lambda = \sqrt{\sum_j p_{\Lambda j}^2}$. This substitution follows from the statistical independence of the Schmidt modes. Because the relative squeezing angles between different modes are random, the momenta add incoherently. This assumption is implicit in the multi-mode nonlinear response derived in refs. 3,39, in which all cross-terms between modes vanish. This treatment is physically justified, as it also ensures gauge invariance of the resulting observables. Meanwhile, the Gaussian statistical distribution for the field momentum should be accordingly changed to be $\chi(P_\Lambda) = \frac{1}{\Gamma(N/2)}P_\Lambda^{N-1}e^{-P_\Lambda^2}$ with $N$ the mode number of the field derived using convolution across different field modes. This leads to the doubly differential ionization rate under an $N$-mode BSV light

$$\Gamma_{\mathrm{QADK}}(p_e, P_\Lambda) \propto \underbrace{\sqrt{2}F_0 P_\Lambda \exp\left\{-\frac{2[2I_p + (p_e - \sqrt{2}A_0 P_\Lambda)^2]^{3/2}}{3\sqrt{2}F_0 P_\Lambda}\right\}}_{\gamma(p_e, P_\Lambda)} \times \underbrace{\frac{1}{\Gamma(N/2)}P_\Lambda^{N-1}e^{-P_\Lambda^2}}_{\chi(P_\Lambda)}. \tag{9}$$

The quantum statistics of the BSV light redistributes the entire weight of the QADK differential rate, which directly manifests as the extended photoelectron kinetic energy spectrum we observe, as visually confirmed in Extended Data Fig. 1. Extended Data Fig. 1a shows the correlated light–electron energy distribution obtained from the QADK theory, in which light energy $E_\Lambda \equiv A_0^2 P_\Lambda^2$ and electron energy $E_e = p_e^2/2$, whereas Extended Data Fig. 1c corresponds to the projection onto the electron energy axis $E_e$. The semiclassical ADK ionization rate represents a single slice of the QADK distribution along a fixed light momentum at $P_\Lambda = 1/\sqrt{2}$ (or $E_\Lambda = A_0^2/2$):

$$\Gamma_{\mathrm{ADK}}(p_e) \propto \Gamma_{\mathrm{QADK}}\left(p_e, P_\Lambda = \frac{1}{\sqrt{2}}\right) \propto F_0 \exp\left\{-\frac{2[2I_p + (p_e - A_0)^2]^{3/2}}{3F_0}\right\}. \tag{10}$$

As the classical slice locates below the dominate region of the positively tilted quantum distribution, two key consequences emerge: (1) the BSV light yields a much higher ionization rate than the coherent light; and (2) the BSV light produces a much higher and broader photoelectron energy distribution.

To quantify the quantum enhancement, we attenuate the BSV field until its peak photoelectron momentum matches that produced by the coherent light, as shown in Extended Data Fig. 1b,d. This condition is met when the average photon number of the BSV light is reduced by a factor of 14 relative to its original value, corresponding to a quantum enhancement factor of 14 based on peak photoelectron momentum matching. The residual discrepancy from the observed 20-fold quantum enhancement might be attributed to quantum correlations across different field modes, which are not included in our current model.

**Analysis of electron number statistics**

Following the development of the QADK theory above, the number of electron $n_e$ can be calculated as

$$n_e = K\gamma(P_\Lambda), \tag{11}$$

where $K$ is a scaling parameter depending on the density of the atom and conversion efficiency, $\gamma(P_\Lambda)$ is the singly differential ionization rate for a specific field momentum $P_\Lambda$:

$$\gamma(P_\Lambda) \propto \sqrt{2}F_0 P_\Lambda \exp\left\{-\frac{2(2I_p)^{3/2}}{3\sqrt{2}F_0 P_\Lambda}\right\}. \tag{12}$$

According to the probability conservation relationship

$$f(n_\mathrm{e})\mathrm{d}n_\mathrm{e} = \chi(P_\Lambda)\mathrm{d}P_\Lambda, \tag{13}$$

the distribution of electron number can be calculated as

$$f(n_\mathrm{e}) = \chi\left[\gamma^{-1}\left(\frac{n_\mathrm{e}}{K}\right)\right]\frac{\mathrm{d}P_\Lambda}{\mathrm{d}n_\mathrm{e}}. \tag{14}$$

The resulting fit to the BSV-driven electron number distribution is presented in Fig. 2a (orange solid line). It is necessary to note that the logic to calculate the electron number distribution is the same as in refs. 3,39, except that we operate in the field momentum space.

### Linear scaling between effective intensity and $g^{(2)}$

The second-order correlation function $g^{(2)}$ is a fundamental quantity characterizing the photon number statistics and quantum properties of light fields. We note that although the rate of a perturbative $n$-photon process generally scales with the $n$th order correlation function $g^{(n)}$ (refs. 70–79), the characterization of the non-classical photon number statistics of the BSV light itself is primarily captured by the standard and widely adopted metric of second-order correlation function $g^{(2)}$. This is because $g^{(2)}$ serves as the primary measure of non-classical statistical properties of the quantum light, directly reflecting the super-Poissonian character inherent to BSV.

For the multi-mode BSV light generated in the present study, the total photon number is distributed across multiple frequency modes[60]. Understanding how $g^{(2)}$ scales with the effective intensity (photon number per mode) is essential for designing quantum light sources with tailored statistical properties. The multi-mode BSV state is generated by applying a multi-mode squeezing operator to the vacuum state, assuming the modes are independent:

$$|\psi\rangle = \bigotimes_{k=1}^{N}\widehat{S}_k(\xi_k)|0\rangle_k, \tag{15}$$

where $N$ is the number of modes[60] and $\widehat{S}_k(\xi_k) = \exp\left(\frac{1}{2}\xi_k^*\hat{a}_k^2 - \frac{1}{2}\xi_k\hat{a}_k^{\dagger 2}\right)$ is the squeezing operator for the $k$th mode, with $\xi_k = r_k\mathrm{e}^{\mathrm{i}\phi_k}$. The total photon number operator is $\hat{n} = \sum_{k=1}^{N}\hat{a}_k^\dagger\hat{a}_k$. The zero-delay second-order correlation function is defined as

$$g^{(2)} = \frac{\langle\hat{n}^2\rangle - \langle\hat{n}\rangle}{\langle\hat{n}\rangle^2}, \tag{16}$$

where $\langle\hat{n}\rangle$ and $\langle\hat{n}^2\rangle$ are the first and second moments of the total photon number distribution.

For a single-mode BSV, the moments are

$$\langle\hat{n}_k\rangle = \sinh^2 r_k, \quad \langle\hat{n}_k^2\rangle = 3\sinh^4 r_k + 2\sinh^2 r_k. \tag{17}$$

For $N$ independent modes, the total moments are

$$\begin{aligned}\langle\hat{n}\rangle &= \sum_{k=1}^{N}\sinh^2 r_k,\\ \langle\hat{n}^2\rangle &= \sum_{k=1}^{N}(3\sinh^4 r_k + 2\sinh^2 r_k) + \sum_{k\neq l}\sinh^2 r_k\sinh^2 r_l\\ &= \sum_{k=1}^{N}(2\sinh^4 r_k + 2\sinh^2 r_k) + \left(\sum_{k=1}^{N}\sinh^2 r_k\right)^2.\end{aligned} \tag{18}$$

Assuming all modes are equally squeezed ($r_k = r$), the moments simplify to

$$\langle\hat{n}\rangle = N\sinh^2 r, \quad \langle\hat{n}^2\rangle = 2N\sinh^4 r + 2N\sinh^2 r + N^2\sinh^4 r. \tag{19}$$

Substituting into $g^{(2)}$, we obtain

$$g^{(2)} = 1 + \frac{2}{N} + \frac{1}{\langle\hat{n}\rangle} \approx 1 + \frac{2}{N} \tag{20}$$

in the limit of large photon number $\langle\hat{n}\rangle$.

The above relationship, $g^{(2)} = 1 + 2/N$, links the second-order correlation function to the effective number of independently amplified modes $N$ (refs. 80,81). It is consistent with known properties of BSV, in which the single-mode limit yields $g^{(2)} = 3$, whereas the expression converges to the coherent state value $g^{(2)} = 1$ as $N \to \infty$. Moreover, it demonstrates that the measured $g^{(2)}$ value of the generated BSV light inherently reflects an effective number of independent, equally squeezed modes. In the case of unequal gain for different modes, it only breaks down to a refinement in the interpretation of $N$ from an exact mode count to the Schmidt number[60], without altering the fundamental conclusions of our analysis. Therefore, we may simply refer to $N$ as the mode number.

For a multi-mode BSV state, the photoelectron momentum spectra are generated through interactions with individual modes and summed across all modes. Consequently, the peak momentum distribution directly reflect the effective intensity per mode, which is determined by the average photon number $\langle\hat{n}_k\rangle$ in each mode:

$$\langle\hat{n}_k\rangle = \frac{\langle\hat{n}\rangle}{N} = \sinh^2 r. \tag{21}$$

The correlation function $g^{(2)}$ can then be directly linked to $\langle\hat{n}_k\rangle$ as

$$g^{(2)} \approx 1 + \frac{2\langle\hat{n}_k\rangle}{\langle\hat{n}\rangle}. \tag{22}$$

In this expression, the average photon number $\langle\hat{n}_k\rangle$ per mode is proportional to the effective intensity $I_\mathrm{eff}$, whereas the total photon number $\langle\hat{n}\rangle$ is proportional to the average pulse energy $P$. The effective intensity scales with the mean photon number per mode, $\langle\hat{n}_k\rangle$, reflecting the fact that the total ionization signal results from the contributions of interactions with each individual field mode. Therefore, a linear scaling between the effective intensity $I_\mathrm{eff}$ and the second-order correlation function $g^{(2)}$ results:

$$I_\mathrm{eff} \propto P[g^{(2)} - 1]. \tag{23}$$

### Effect of pulse duration

In our experiments, although the pulse durations of the 70-fs coherent pulse and 150-fs BSV pulse are different, they have minor influence on our comparison for the two key reasons.

First, in our measurement, the comparison between coherent and BSV light sources focuses on the effective peak intensity. Tunnelling ionization, as a highly nonlinear process, is predominantly driven by the peak electric field of the pulse. Our angular streaking technique directly measures this effective peak intensity, enabling a controlled comparison at the intensity level relevant to the process.

Second, to confirm that the observed photoelectron kinetic energy spectrum is insensitive to pulse duration, we performed a simulation using coherent pulses of varying durations. As shown in Extended Data Fig. 2, the photoelectron kinetic energy spectra from tunnelling ionization driven by 70 fs and 150 fs coherent pulses are virtually identical. This demonstrates that over this duration range, the temporal envelope does not introduce qualitatively different physics that could account for the observed quantum enhancement.

Therefore, although identical pulse durations for coherent and BSV pulses are, in principle, ideal for comparison, the evidence supports that our comparison is physically sound, and the conclusions about quantum enhancement remain robust.

### Choice of elliptical polarization

The choice of elliptical polarization for the driving field, as opposed to linear, was a deliberate and important aspect of our experimental design. In linear polarization, the photoelectron momentum distribution is affected by rescattering and interference effects, resulting in a complex structure that cannot be uniquely mapped to a single value of the vector potential amplitude. The elliptical streaking field, by contrast, enables angular streaking, which generates a photoelectron momentum distribution that is centred around an ellipse corresponding to the vector potential amplitude.

The use of elliptical polarization was further necessitated by signal-to-noise considerations. Although circular polarization provides an ideal rotating field for angular streaking, it distributes pulse energy equally between two orthogonal axes, effectively halving the peak intensity along any direction. Given the experimental challenge of generating intense BSV, circular polarization would have resulted in ionization yields too low for statistically robust coincidence measurements. Elliptical polarization preserves a significant rotating field component for angular streaking while maintaining a strong field amplitude along the major axis, thereby ensuring sufficient ionization rates without compromising the physical interpretation of the streaking signal. The selected ellipticity thus represents an optimal trade-off between streaking interpretation and ionization efficiency, in which the peak momentum along the direction of maximum emission probability sits at $\varepsilon A_0$.

### Calibration of effective intensity

In both the QADK and semiclassical ADK theories, electron momentum, rather than kinetic energy, is directly involved as shown in equations (8) and (10). Therefore, in our study, we calibrate the intensity using photoelectron momentum distributions rather than energy. As shown in Extended Data Fig. 3, these distributions maintain a Gaussian-like profile across all $g^{(2)}$ values, enabling a reliable and accurate extraction of the peak momentum. This photoelectron peak momentum directly reflects the effective intensity of the BSV field. Here, the enhanced tails of the momentum distribution are a demonstration of the quantum statistical nature of the BSV light.

Nevertheless, in Fig. 3a of the main text, we present the results in terms of kinetic energy in eV rather than momentum, because the energy unit of eV is a more intuitive and comprehensible unit, leading to an easier comparison with existing literature[40]. It should be noted that the calibrated value of the effective intensity differs slightly depending on whether the matching is performed using kinetic energy or momentum, as $\langle p_e^2\rangle \neq \langle p_e\rangle^2$. That said, the key conclusions of our work, particularly the quantum enhancement and linear scaling of the effective intensity with $g^{(2)}$, remain invariant when we carry out the matching using the same physical observable, be it momentum or kinetic energy.

### Mode analysis of photon number statistics

Through singular value decomposition, a multi-mode field can be expressed on the basis of independent effective modes. The effective number of these modes, generally quantified by the Schmidt number[60], is a measure of the number of independent, uncorrelated components within the BSV field, as widely used in both theory and experiments. In the present study, the generated BSV light contains a single spatial mode and multiple frequency modes. To confirm this, we performed spectral filtering of the BSV light using a standard 4-$f$ monochromator. The slit in the monochromator can change the spectral bandwidth and thus alter the frequency mode number of the BSV source. Extended Data Fig. 4a demonstrates the measured second-order correlation function $g^{(2)}$ of the BSV light after the 4-$f$ monochromator as a function of the slit width in the 4-$f$ monochromator, which manipulate the spectral bandwidth and the frequency mode of the BSV light. After spectral filtering, we achieved $g^{(2)}$ values up to 2.6, which confirms the contribution of multiple frequency modes to our BSV source. We note that, to maintain the high peak intensity required for atomic ionization in our experiment, we did not use spectral filtering to the generated BSV that enters the vacuum chamber for the ionization of the atoms as a multi-mode light.

In our experiments, we primarily control the $g^{(2)}$ parameter of the BSV source by varying the pump power incident on the nonlinear crystal. Extended Data Fig. 4b shows the measured dependence of $g^{(2)}$ on pump power, demonstrating that we can systematically tune this parameter. To validate the independent, equally squeezed multi-mode characterization of the BSV light, we measured photon number distributions at distinct $g^{(2)}$ values. Extended Data Fig. 5 presents six typical photon number distributions (blue shades), in which lower $g^{(2)}$ values exhibit multi-mode characteristics: the peak is shifted towards higher photon numbers, and the overall distribution shows a progressively closer resemblance to a Poissonian distribution. For quantitative comparison, we calculated the expected photon number distributions, shown in Extended Data Fig. 5 (blue curves), according to the multi-mode photon number distribution[3]:

$$P_{\mathrm{BSV}}(n,N)=\frac{n^{N/2-1}}{\Gamma(N/2)}\left(\frac{N}{2\langle n\rangle}\right)^{N/2}\mathrm{e}^{-\frac{Nn}{2\langle n\rangle}}, \quad (24)$$

where $\Gamma$ is the gamma function, $N$ is the number of modes, and $n$ denotes the photon number. The effective mode number is estimated as $N = 2/[g^{(2)}-1]$ (equation (20)). The good agreement between measurement and calculation provides strong evidence for the multi-mode characterization of our BSV source and the validity of equation (20) derived under the assumption of multiple independent equally squeezed modes. This treatment is a common practice for characterizing sources based on parametric down-conversion and is consistent with the methodology used in established works[1,3,60].

## Data availability

The data supporting this paper are available from the corresponding authors upon reasonable request.

**Acknowledgements** We thank J.-W. Pan for the discussions. This work was supported by the Quantum Science and Technology—National Science and Technology Major Project (grant no.

2024ZD0300700), the National Natural Science Foundation of China (grant nos. 12521003, 12241407, 12474341 and 12574371) and the Shanghai Pilot Program for Basic Research (grant no. TQ20240204).

**Author contributions** J.W. conceptualized the study. Z.J., S.P., J.C., C.Z., Y.W., L.H., D.W., W.L. and J.W. carried out the experimental studies. M.Z., R.Z., J.L., S.X., S.J. and H.N. carried out the theoretical studies. All authors participated in the discussions and contributed to the paper.

**Competing interests** The authors declare no competing interests.

**Additional information**

**Supplementary information** The online version contains supplementary material available at https://doi.org/10.1038/s41586-026-10485-9.

**Correspondence and requests for materials** should be addressed to Shicheng Jiang, Hongcheng Ni or Jian Wu.

**Peer review information** *Nature* thanks Jeehwan Kim, Marco Lisker and the other, anonymous, reviewer(s) for their contribution to the peer review of this work. Peer reviewer reports are available.

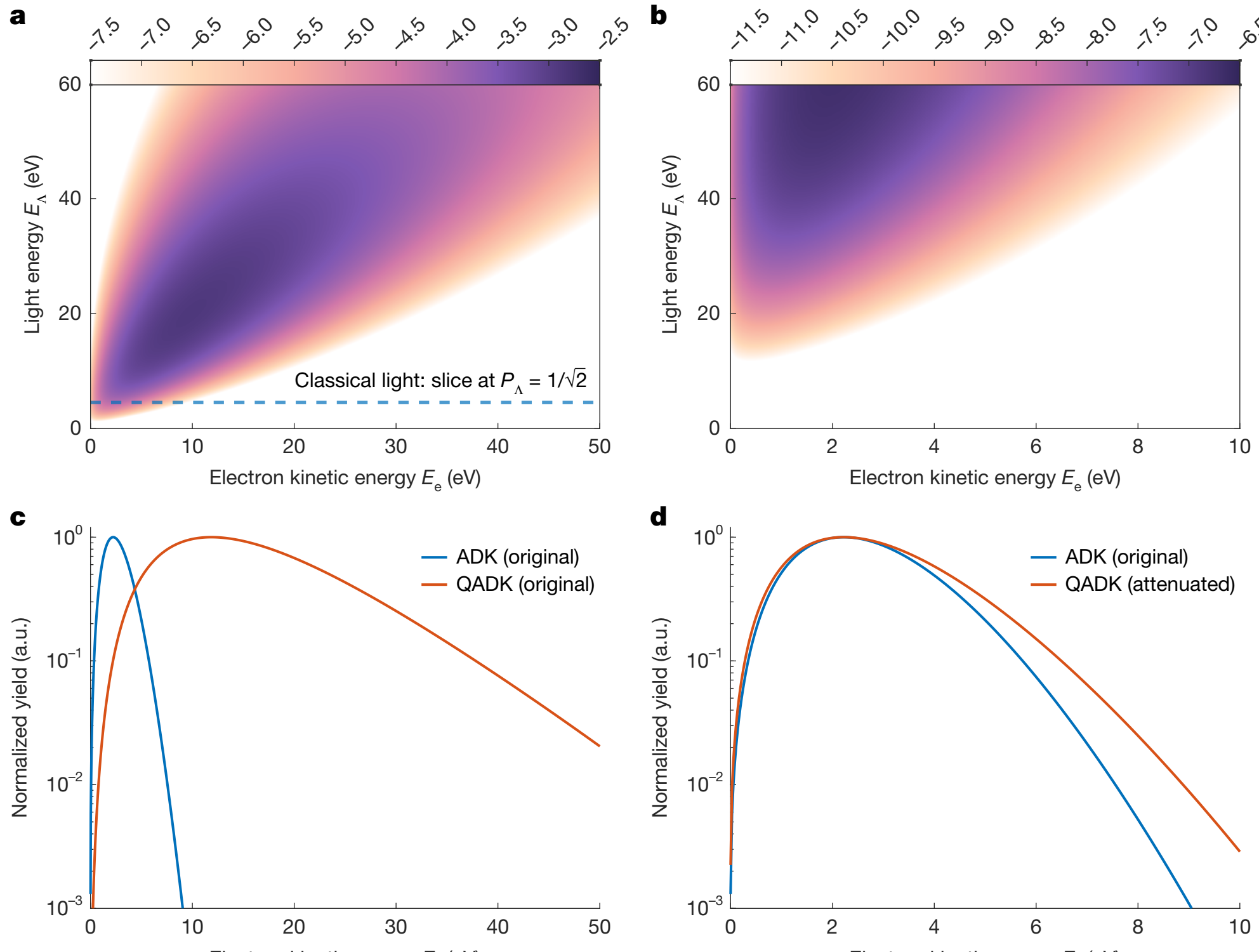

**Extended Data Fig. 1 | Simulated quantum-boosted photoelectron spectra.** **a**,**c**, Correlated light–electron energy distribution (**a**, logarithmic scale) and its projection onto the electron energy axis $E_e$ (**c**, logarithmic scale, normalized). Identical average photon number for coherent and BSV lights are used. **b**,**d**, Correlated light–electron energy distribution (**b**, logarithmic scale) and its projection onto the electron energy axis $E_e$ (**d**, logarithmic scale, normalized). The BSV light is attenuated by a factor of 14 so that its peak momentum matches that of the coherent light of the original intensity.

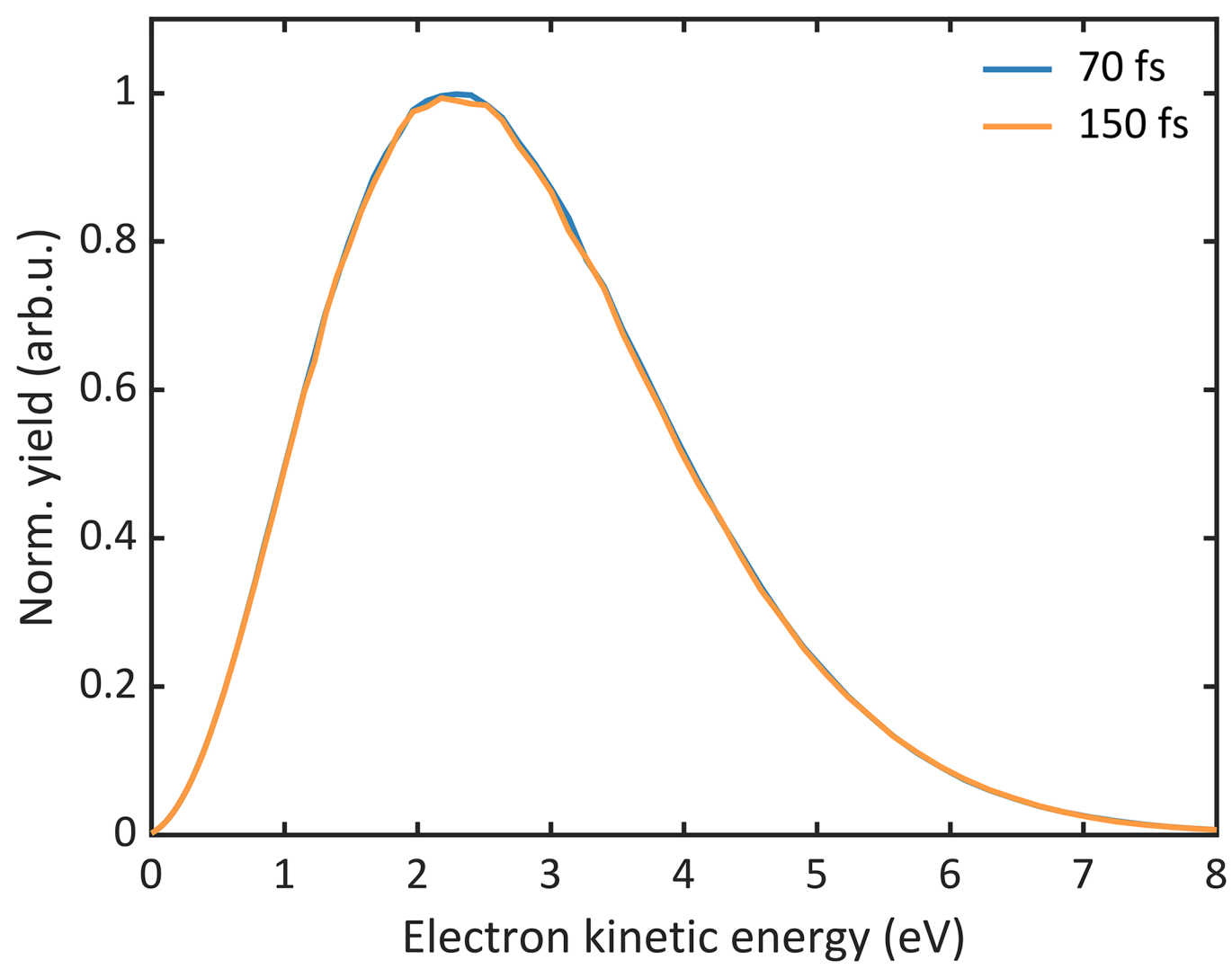

**Extended Data Fig. 2 | Simulated photoelectron spectra under coherent lights with different pulse durations.** Blue and yellow curves denote the electron kinetic energy spectrum simulated with the pulse duration of 70 fs and 150 fs, respectively.

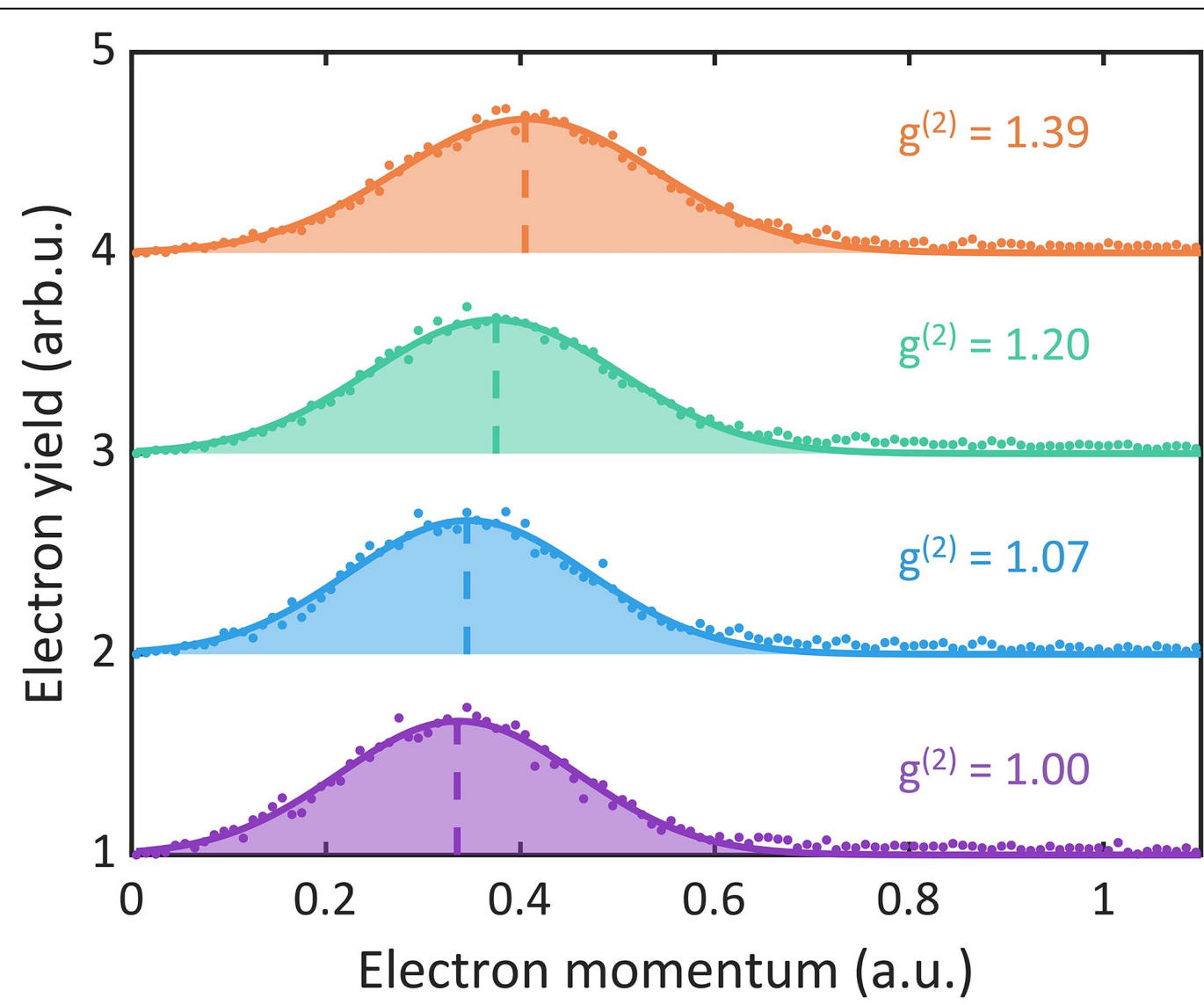

**Extended Data Fig. 3 | Photoelectron momentum spectra resulting from BSV light with varying second-order correlation function $g^{(2)}$.**

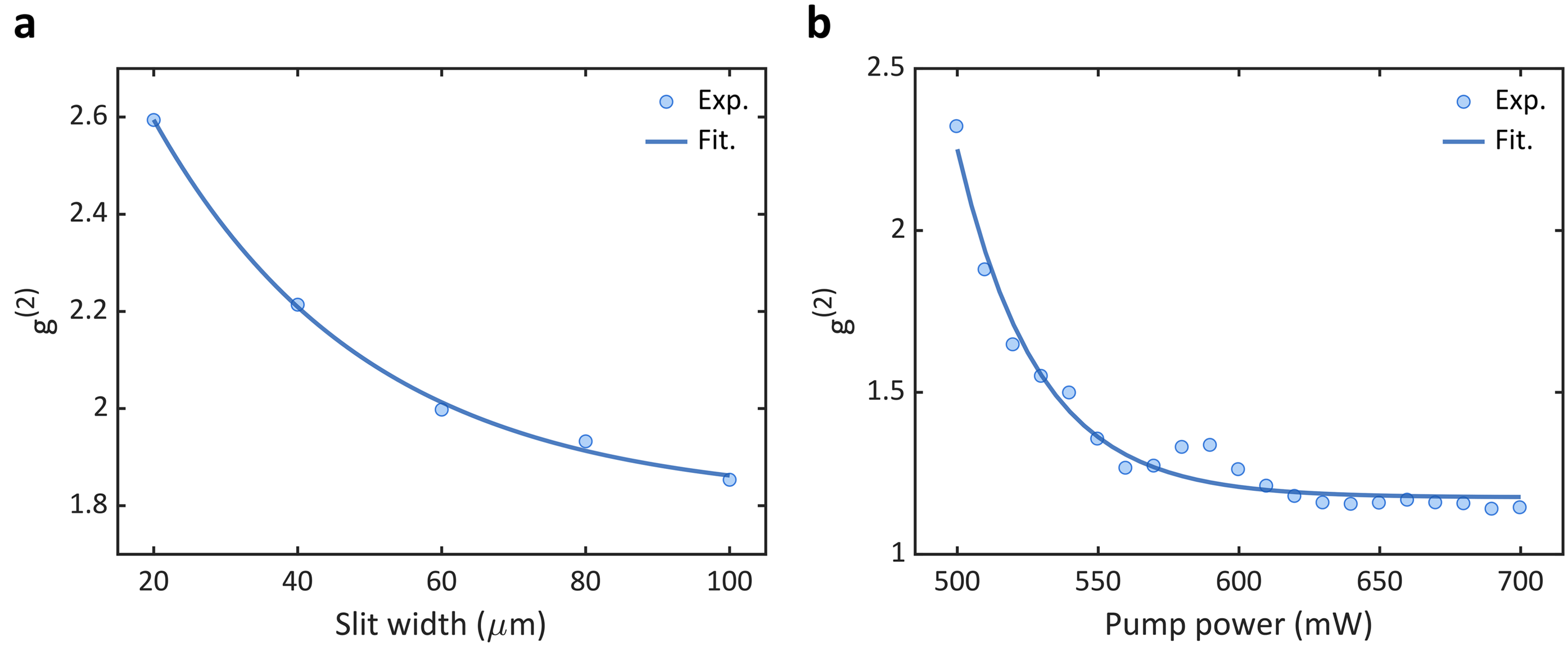

**Extended Data Fig. 4 | Measured second-order correlation function $g^{(2)}$ of the generated BSV light.** (**a**) Second-order correlation function as a function of the slit width in the 4-$f$ monochromator. (**b**) Second-order correlation function as a function of pump power of the pulse entering the nonlinear crystal.

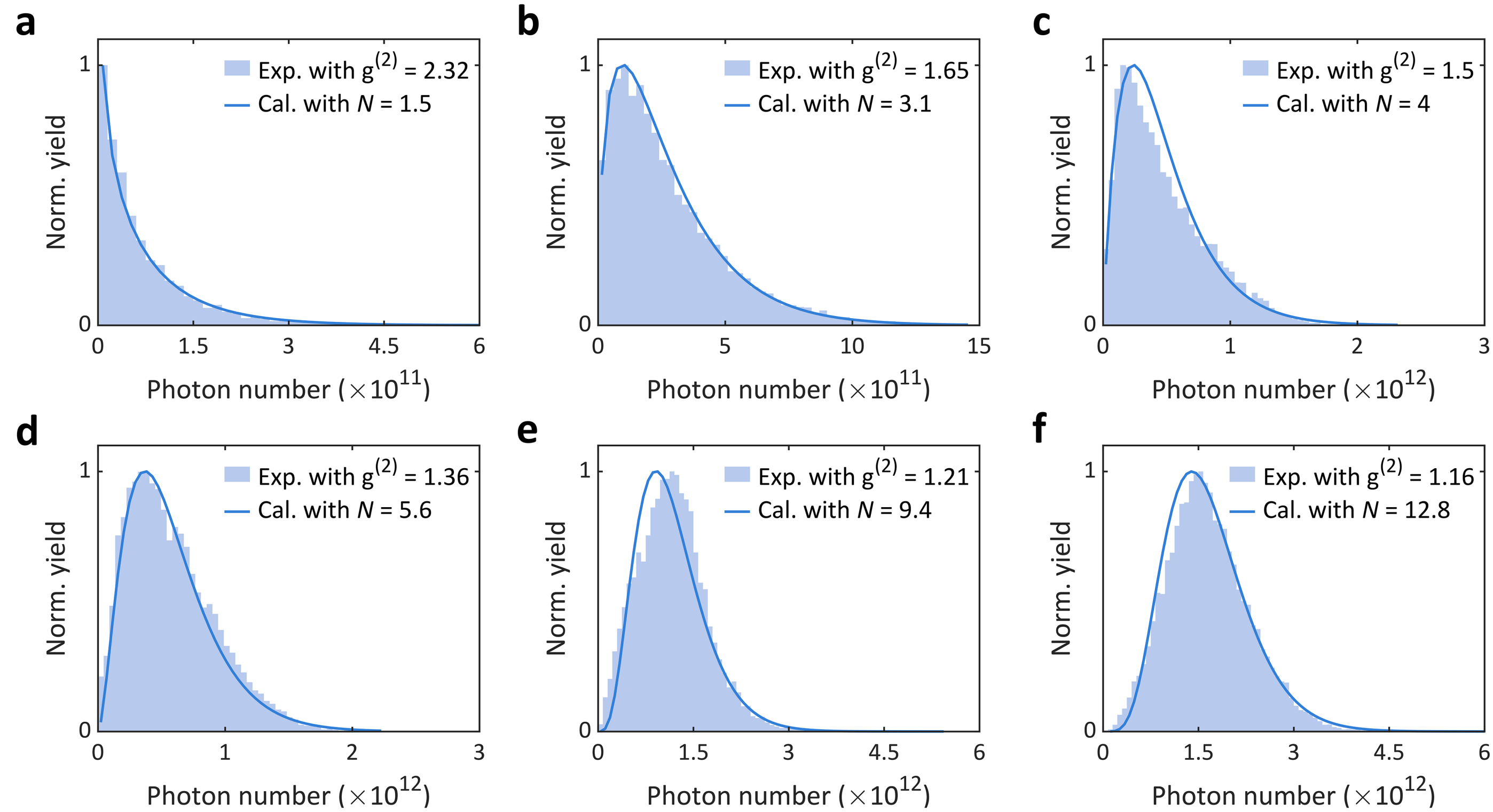

**Extended Data Fig. 5 | Photon number distributions of the generated BSV light with varying $g^{(2)}$ parameters.** Blue shades denote the photon number distributions measured in experiments, and blue curves are calculated according to the multi-mode formula only using the measured average photon number and effective mode number estimated from $N = 2/[g^{(2)} - 1]$.